# Ligand Engineering for Precise Control of Ultrathin $CsPbI_3$ Nanoplatelet Superlattices for Efficient Light-Emitting Diodes

Jongbeom Kim[1,†], Woo Hyeon Jeong[2,3,†], Junzhi Ye[3,4]*, Allison Nicole Arber[5], Vikram[5], Donghan Kim[1], Yi-Teng Huang[3,6], Yixin Wang[3], Dongeun Kim[1], Dongryeol Lee[1], Chia-Yu Chang[3], Xinyu Shen[7], Sung Yong Bae[3], Ashish Gaurav[3], Akshay Rao[8], Henry J. Snaith[7], M. Saiful Islam[5]*, Bo Ram Lee[2]*, Myoung Hoon Song[1]* Robert L. Z. Hoye[3]*

[1]Department of Materials Science and Engineering, Ulsan National Institute of Science and Technology (UNIST), Ulsan, 44919 Republic of Korea

[2]School of Advanced Materials Science and Engineering, Sungkyunkwan University, Suwon, 16419, Republic of Korea

[3]Inorganic Chemistry Laboratory, University of Oxford, Oxford, OX1 3QR, United Kingdom

[4]Institute of Polymer Optoelectronic Materials and Devices, Guangdong Basic Research Center of Excellence for Energy & Information Polymer Materials, State Key Laboratory of Luminescent Materials and Devices, School of Materials Science and Engineering, South China University of Technology, Guangzhou 510640, China.

[5]Department of Materials, University of Oxford, Oxford, OX1 3PH, United Kingdom

[6]Graduate Institute of Photonics and Optoelectronics and Department of Electrical Engineering, National Taiwan University, Taipei 10617, Taiwan

[7]Clarendon Laboratory, Department of Physics, University of Oxford, Oxford, OX1 3PU United Kingdom

[8]Cavendish Laboratory, University of Cambridge, Cambridge CB3 0US, United Kingdom

[†]These authors contributed equally.

*Correspondence: junzhi.ye@chem.ox.ac.uk (J.Y.), saiful.islam@materials.ox.ac.uk (M.S.I.), brlee@skku.edu (B.R.L.), mhsong@unist.ac.kr (M.H.S.), robert.hoye@chem.ox.ac.uk (R.L.Z.H.)

## ABSTRACT

Strongly-confined perovskite nanoplatelets (PeNPLs) offer opportunities not found in conventional isotropic nanocubes, especially in producing linearly polarized light, as well as enhancing outcoupling through control over the transition dipole moment. But this requires ultrathin nanoplatelets with three or fewer monolayers of $PbI_6$ octahedra across the thickness, which are challenging to synthesise uniformly, and their luminescence is strongly affected by surface defects. Together, these limit the performance of ultrathin PeNPLs in light-emitting diodes (LEDs). Here, we address these challenges with an ancillary ligand engineering strategy. We demonstrate that ligands with phosphoryl functional groups strongly bind to the perovskite surface, while having an organic backbone that is not sterically bulky ensures high ligand density. By modulating nucleation and growth, these ancillary ligands lead to monodisperse PeNPLs that stack more uniformly when self-assembled into superlattices, with suppressed agglomeration. As a result, from edge-up PeNPL superlattices, we achieve enhanced degree of polarization, while from face-down PeNPL superlattices, we achieve enhanced outcoupling that results in LEDs with 13.1% external quantum efficiency, the highest reported for ultrathin PeNPL LEDs. This work establishes ancillary ligand-induced synthesis as a decisive route to achieve uniform nanoplatelets with robust orientation control, enabling full utilization of the multifunctionality of anisotropic PeNPLs.

## INTRODUCTION

Colloidal semiconductor nanocrystals (NCs) offer advantages for optics and photonics, as well as for fundamental investigations of light-matter properties[1, 2, 3]. To fully realize these benefits, it is critical to achieve highly monodisperse NCs, which requires precise control over nucleation and growth via well-regulated reaction conditions[4, 5, 6, 7, 8]. Lead halide perovskites, with the general formula $APbX_3$ (A = formamidinium, methylammonium, Cs; X = Cl, Br, I), have emerged as one of the most promising NC materials, due to their narrow full-width at half-maximum (FWHM), defect tolerance, and the ability to achieve high luminescence quantum yields with cost-effective solution processing methods[9, 10, 11, 12, 13]. In particular, perovskite nanoplatelets (PeNPLs), classified by the number of monolayers (MLs) of lead-halide octahedra across the thickness, can have their emission wavelength tuned not only by changing their composition, but also by tuning their thickness. Thinner PeNPLs exhibit a blue-shift in their emission due to stronger quantum and dielectric confinement effects across the out-of-plane direction[14, 15]. Furthermore, the anisotropy in their confinement leads to exciton fine structure splitting, which can enable linearly polarized light emission[16, 17, 18].

Prior studies have shown that the pronounced linearly polarized emission from PeNPLs is strongly thickness-dependent and most reliably obtained in the ultrathin regime (typically $n \leq 3$, where $n$ is the number of monolayers of Pb-halide octahedra across the thickness of the PeNPL). In ultrathin PeNPLs, anisotropic confinement maximizes the fine structure splitting in the bright excitonic manifold, where the excitonic state associated with the out-of-plane direction is raised in energy more than the doubly degenerate in-plane excitonic states [18, 19, 20]. Thicker nanoplatelets generally exhibit a weaker polarization response due to reduced exciton fine structure splitting. Beyond the intrinsic excitonic effects, the pronounced shape anisotropy of ultrathin PeNPLs can also promote orientational ordering and superlattice assembly, allowing control of the transition dipole moment in films[21]. Therefore, achieving strong linear polarization and enabling ordered assembly requires precise and uniform thickness control in the ultrathin regime, since even a small fraction of non-uniformity by having a mixture of different ML PeNPLs can rapidly reduce the multifunctionality of the assembled films. However, attaining this uniformity among nanoplatelets is highly challenging, given their surface-dominated nature. The high surface area-to-volume ratio amplifies sensitivity to ligand dynamics and surface defects, thereby requiring robust surface passivation.

PeNPLs present inherent challenges due to their fast reaction kinetics, which stems from their ionic nature and low formation energy, making it difficult to achieve nanoplatelets that are uniform in thickness[5, 22, 23]. This results in a distribution of emission wavelengths that compromise their optical properties, including their colour saturation (required for ultrahigh definition displays) and degree of linear polarization (required for optical communications, 3D displays and bioimaging)[24, 25]. Such limitations represent a fundamental bottleneck across all classes of NPLs, not just halide perovskites, since achieving colloidal uniformity and structural stability are essential for fully exploiting their optoelectronic potential[26].

For the synthesis and dispersion of PeNPLs, long chain organic ligands such as oleate ($OA^-$) and oleylammonium ($OAm^+$) are commonly used[27, 28, 29]. While these native ligands play a crucial role in PeNPL synthesis, their weak surface binding affinity leads to dynamic ligand desorption, resulting in poor colloidal stability and surface defect formation, including halide vacancies and uncoordinated Pb ions[30, 31, 32]. These defects can lead to aggregation between PeNPLs or decomposition, resulting in a distribution in PeNPL thicknesses, which prevents the realization of colour-pure LEDs with a high degree of linear polarisation. This instability is further exacerbated during thin film formation, due to ligand desorption, diminishing both the stability and optical properties of PeNPLs[33, 34]. Moreover, these insulating ligands severely hinder charge-carrier transport and injection, making them unsuitable for high-performance LED applications[32, 35]. Therefore, to achieve efficient red-emitting iodide-based PeNPLs, it is essential to develop monodisperse PeNPLs while implementing advanced passivation strategies to overcome the limitations posed by native ligands.

Motivated by these challenges, we systematically investigate an ancillary ligand-induced synthesis approach to precisely control the nucleation and growth of $CsPbI_3$ PeNPLs. This involves introducing additional Lewis base ligands (*i.e.*, ancillary ligands) to bind to the perovskite surface in addition to the oleate and oleylammonium ligands used in the synthesis of pristine PeNPLs. We herein find that the key determinants of an effective ancillary ligand are its backbone structure and Pb-coordination strength with the perovskite surface. Ancillary ligands that optimally satisfy these two determining factors modulate the crystal growth kinetics, enabling the formation of uniform 3ML-thick PeNPLs while simultaneously passivating surface defects. These PeNPLs not only retain homogeneous emission upon thin-film formation but also self-assemble into well-ordered superlattices with significantly enhanced optical properties compared to pristine PeNPLs, owing to their well-passivated surfaces that suppress agglomeration. By achieving fine control over the orientation of the NPLs in superlattices, we improve the degree of polarization (DOP) in edge-up NPLs.

Conversely, face-down oriented PeNPL superlattices enhance outcoupling in the direction normal to the substrate in LEDs, leading to high external quantum efficiencies (EQEs) reaching 13.1%, which represents the highest value reported for LEDs based on ultrathin PeNPLs (*i.e.*, 3 ML or lower), where exciton fine structure splitting becomes sufficient for realising polarised light emission. Taken together, our results establish ancillary ligand-induced synthesis of uniform-ML PeNPLs as a robust route to fully exploit the multifunctionality of these anisotropic nanocrystals for next-generation photonics and displays.

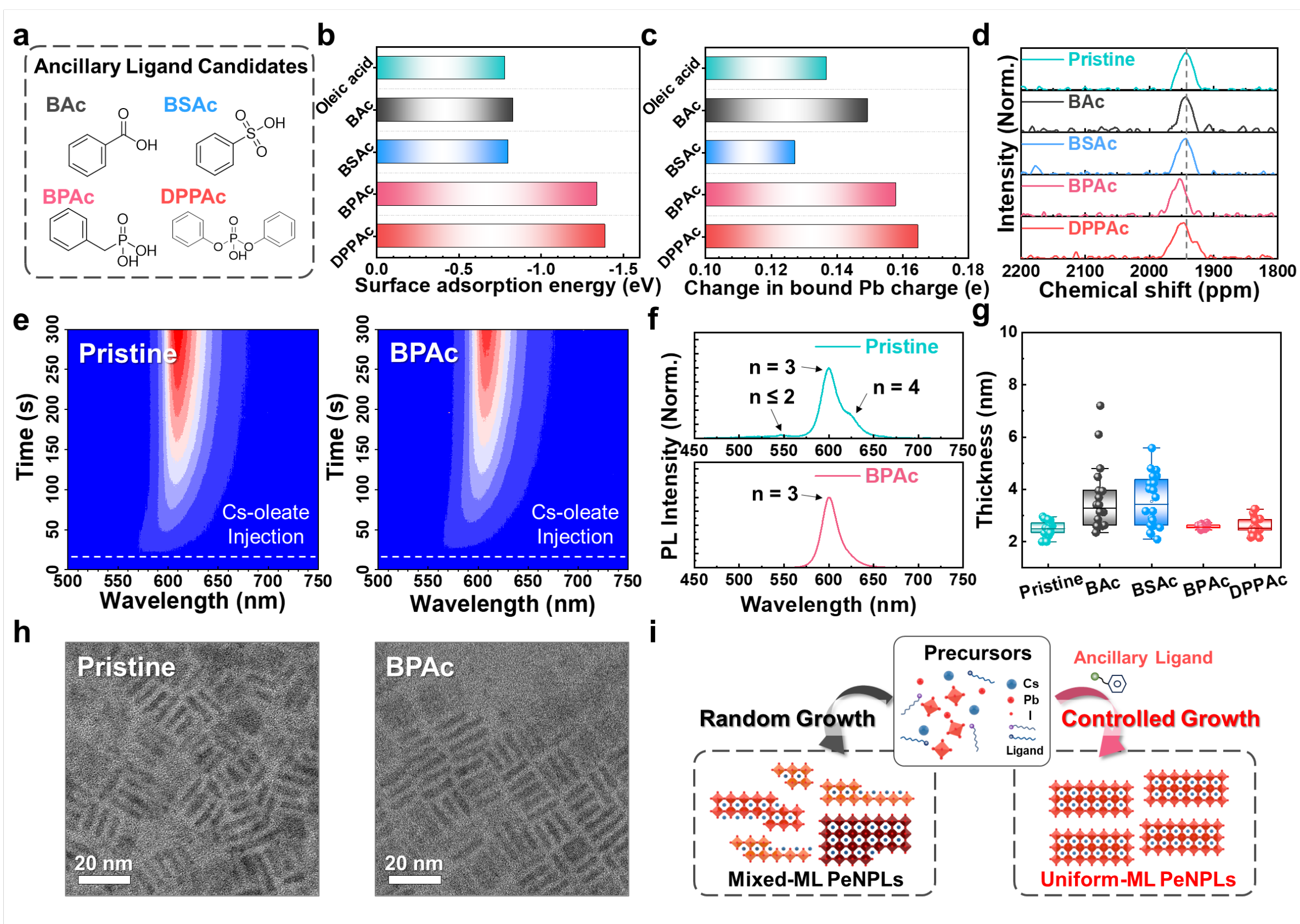

**Fig. 1 | Ancillary ligand engineering to achieve well-ordered, monodisperse $CsPbI_3$ PeNPLs**

**a** Candidate ancillary ligands considered in this work: benzoic acid (BAc), benzene sulfonic acid (BSAc), benzyl phosphonic acid (BPAc), diphenyl phosphate (DPPAc). **b** Calculated surface adsorption energy of ligands on the perovskite surface. **c** Calculated Bader charge on the surface Pb atom bound to the ligand molecule referenced to the charge of an unbound surface Pb atom. These charges increase upon binding, which correlates to the deshielding of Pb observed by nuclear magnetic resonance (NMR). **d** Liquid-phase $^{207}$Pb NMR spectra of $PbI_2$ precursor solutions without and with ancillary ligands. **e** *In-situ* photoluminescence spectra recorded during synthesis under ambient air conditions, excited with a 405 nm CW laser (37.3 mW $cm^{-2}$). **f** Photoluminescence spectra of pristine-PeNPL and BPAc-PeNPL colloidal solutions. **g** Thickness distribution histogram of $CsPbI_3$ PeNPLs determined from TEM measurements from Fig. 1h and Supplementary Fig. 8. The target thickness during synthesis was 3 monolayers. **h** High-resolution transmission electron microscopy (TEM) images of edge-up oriented pristine-PeNPLs and BPAc-PeNPLs. These nanoplatelets were dispersed in a hexane solvent and drop-cast onto a Cu TEM grid. **i** Schematic illustration of conventional synthesis and ancillary ligand-induced synthesis for PeNPLs.

# RESULTS

## Controlling ligand-perovskite surface interactions to achieve uniform PeNPLs

We hypothesized that the introduction of ligands strongly coordinating to the perovskite precursor monomers in colloidal solution can retard PeNPL formation, thus enabling regulation of crystallization and greater control over the final thickness of the PeNPLs. Within this hypothesis, the functional group of the ligand would play a decisive role, as would the organic backbone of the ligands, which can influence the growth kinetics through steric hindrance[38]. Guided by this hypothesis, we compared ligands with a carboxylic acid, sulfonic acid and phosphonic acid functional group (keeping the organic backbone a benzyl group), *i.e.*, benzoic acid (BAc), benzene sulfonic acid (BSAc) and benzyl phosphonic acid (BPAc). We selected these functional groups for comparison, since they have oxo-groups available for coordination with $Pb^{2+}$ as Lewis bases. We also compared ligands with the same phosphoryl functional group, but changed the organic backbone, *i.e.*, BPAc *vs.* diphenyl phosphate (DPPAc). The ancillary ligand candidates are illustrated in **Fig. 1a**.

Understanding the interactions between Pb precursors and coordinating ligands at the atomic scale is critical for elucidating how crystallization can be regulated. Prior to experimental investigation, density functional theory (DFT) calculations were performed to provide atomistic insights into ligand-surface interactions to clarify the role of specific ligands (Computational details in **Methods**).

The calculations show that DPPAc and BPAc have the highest surface adsorption energies (**Fig. 1b**) on the (001)-terminated surface of the perovskite. Surface adsorption energy is defined as the energy reduction upon ligand adsorption to the perovskite surface after structural optimization of the complex. The strength of adsorption or binding can be related to the degree of $PbI_6$ octahedral di stortion simulated at the surface, as well as the length of the bond formed between the surface a nd the ligand molecules (**Supplementary Fig. 1**). DPPAc and BPAc were found to induce the hig hest degree of structural distortion, and both have the shortest Pb–O bond length of 2.46 Å, com pared to 2.63 Å for oleic acid, the weakest-bound ligand. The calculated projected density of states (PDOS) for Pb and O atoms involved in bonding between BPAc and the perovskite surface show significant Pb-O hybridisation in the occupied (valence) region, indicating strong covalency between the ligand and perovskite surface (**Supplementary Fig. 2**). This is consistent with O from the phosphoryl functional group in BPAc acting as a Lewis base and coordinating with the empty Pb 6p states, with electron density shared between them. Bader charge analysis (**Fig. 1c**) assigns slightly more electron density to O than to Pb in the Pb-O bond because of the higher electronegativity of the O atom. Comparing all ligands, the DPPAc and BPAc ligands exhibited the largest polarization of the Pb-ligand bond (*i.e.*, largest change in bound Pb charge, **Fig. 1c**), which is consistent with the phosphoryl group being the functional group binds strongest to the perovskite surface.

To further substantiate differences in ligand-precursor interactions, we performed liquid-phase $^{207}$Pb nuclear magnetic resonance (NMR) measurements of the $PbI_2$ precursor solutions with and without the ancillary ligand added. The $^{207}$Pb NMR chemical shift is highly sensitive to the coordination environment of $Pb^{2+}$ [40]. As shown in **Fig. 1d**, the precursor solutions containing BPAc and DPPAc exhibit a clear chemical shift from 1943.4 ppm (pristine) to 1953.2 ppm and 1949.7 ppm, indicating strong coordination of BPAc and DPPAc to $Pb^{2+}$. That these ligands have the largest change in chemical shift substantiates the simulated surface adsorption energies (**Fig 1b**) showing these ligands to have the strongest binding to the perovskite surface via the P=O groups.

Based on these atomic-scale insights, we synthesized colloidal $CsPbI_3$ PeNPLs using the LARP method (detailed in **Methods**). Five types of samples were prepared for comparison: pristine-PeNPLs synthesized without ancillary ligands (pristine-PeNPLs), and PeNPLs with ancillary ligands incorporated (BAc-PeNPLs, BSAc-PeNPLs, BPAc-PeNPLs and DPPAc-PeNPLs), prepared by adding the ligands to the $PbI_2$ precursor solution (2.9 mM concentration). To elucidate the mechanism by which ancillary ligands regulate PeNPL growth, we conducted *in-situ* photoluminescence (PL) measurements during synthesis. A 405 nm wavelength continuous-wave (CW) laser was used as the excitation source, and the PL spectra were recorded at 1 s time intervals following the injection of the Cs-oleate precursor into the $PbI_2$-ligand solution. As shown in **Fig. 1e** and **Supplementary Fig. 3-4**, pristine-PeNPLs exhibited a rapid onset of PL signal followed by a gradual redshift in emission wavelength, indicating the formation of PeNPLs with a mixture of different thicknesses, arising through uncontrolled nucleation and growth. BAc- and BSAc-PeNPLs developed PL emission faster, reaching saturation within 250 s before declining, indicating that they did not retard crystal growth and instead destabilized the reaction dynamics. In contrast, BPAc- and DPPAc-PeNPLs displayed a delayed PL onset and a slower spectral shift, suggesting suppressed nucleation and more regulated crystal growth dynamics. This trend is quantified in **Supplementary Fig. 5**, which plots the evolution of the PL peak wavelength as a function of reaction time. Pristine-, BAc- and BSAc-PeNPLs showed a fast redshift from ~580 to 609 nm wavelength, while BPAc- and DPPAc-PeNPLs gradually evolved from 570 to 608 nm wavelength, indicating more controlled lateral and vertical growth. Additionally, the PL intensity profiles during the early reaction stages (**Supplementary Fig. 6**) reveal that BPAc- and DPPAc-PeNPLs exhibit a slower rise in PL intensity compared to other samples. These observations suggest that BPAc and DPPAc ancillary ligand slow down nucleation and the subsequent crystal growth kinetics. Such observations are consistent with these ligands binding more strongly with $Pb^{2+}$, since they could delay the availability of $Pb^{2+}$ for perovskite formation.

**Fig. 1f** and **Supplementary Fig. 7** show the ultraviolet-visible absorption and PL spectra of the PeNPL dispersions. Given that the out-of-plane dimensions of the PeNPLs are smaller than the Bohr diameter of $CsPbI_3$ (~12 nm [41]), the optical properties of these nanocrystals are strongly influenced by quantum confinement effects. The absorption spectra of all PeNPL samples exhibit a pronounced excitonic peak, directly confirming strong quantum confinement. Pristine-PeNPLs exhibit multiple PL peaks centered at 550, 600, and 627 nm wavelength (**Fig. 1f**), corresponding to nanoplatelets with $n$ = 2, 3, and 4 [14], respectively. These multiple emission peaks indicate poor thickness uniformity, which is a known challenge in PeNPL systems, given that these PeNPLs are achieved under kinetically controlled growth conditions, rather than thermodynamic growth conditions. Multi-emission peaks were also found in the BAc-, BSAc- and DPPAc-PeNPLs, even peaks centered at 680 nm wavelength, which corresponds to nanoplatelets with $n \geq 5$ MLs. By contrast, BPAc-PeNPLs exhibit a single, sharp emission peak at 600 nm wavelength (FWHM ~ 21 nm), consistent with the formation of uniform 3ML-thick nanoplatelets. The absorption spectra further corroborate this observation, with the absence of higher or lower energy transitions confirming the suppression of undesired thickness variants.

To further validate the morphological and structural uniformity, high-resolution transmission electron microscopy (HR-TEM) was performed on edge-up oriented PeNPLs (**Fig. 1g-h** and **Supplementary Fig. 8**) and face-down oriented PeNPLs (**Supplementary Figs. 9-10**). Detailed HR-TEM analysis revealed the thicknesses of 2.50 ± 0.29 nm for pristine-PeNPLs, 3.63 ± 0.12 nm for BAc-PeNPLs, 3.54 ± 0.98 nm for BSAc-PeNPLs, 2.57 ± 0.06 nm for BPAc-PeNPLs, and 2.64 ± 0.39 nm for DPPAc-PeNPLs. Notably, among all samples, BPAc-PeNPLs is the only one that shows a narrow thickness distribution near the target 3 ML thickness of $CsPbI_3$ PeNPLs (≈2.6 nm)[18, 25, 42]. PeNPLs with this ancillary ligand also had aspect ratios that were closest to unity (**Supplementary Fig. 10**). The significantly narrower thickness distribution in BPAc-PeNPLs directly reflects enhanced monodispersity, consistent with the optical results. However, it is curious that BPAc led to improved monodispersity than DPPAc, despite both having P=O groups that bind to $Pb^{2+}$, and DPPAc and BPAc having similar surface adsorption energies. To understand this apparent discrepancy and rationalize the effects of steric hindrance during NPL growth, we conducted $^1$H-NMR to characterize the surface chemistry of BPAc- and DPPAc-PeNPLs (**Supplementary Fig. 11**). Ferrocene (0.5 mg $mL^{-1}$) was introduced into each PeNPL solution in DMSO-$d_6$ as an internal reference for quantitative analysis. In the $^1$H-NMR spectra, the resonance at 4.17 ppm is assigned to ferrocene, and the signal at 5.33 ppm corresponds to the vinylic protons (-HC=CH-) of the native ligands[43, 44]. Quantitative integration of these peaks relative to ferrocene shows that DPPAc-PeNPLs have significantly decreased amounts of native oleate/oleylammonium ligands than pristine- and BPAc-PeNPLs. This difference can be rationalized from the more sterically bulky backbone of DPPAc, which would hinder the attachment of native ligands on the perovskite surface during growth and purification, leading to reduced colloidal stability and increased polydispersity. Overall, the results in **Fig. 1** underscore two notable conclusions. First, the incorporation of functional groups such as P=O, which exhibit strong binding to Pb, is crucial for regulating crystal growth, thereby enabling the synthesis of uniform-ML PeNPLs. Second, even when crystal growth is effectively controlled, the steric bulk of the ancillary ligands can modulate the surface chemistry and, in turn, alter the uniformity in thickness of the resulting PeNPLs.

Based on the understanding of the nucleation and growth kinetics of ancillary ligand incorporation, to investigate the impact of the BPAc ligands on the optical properties of PeNPLs, we performed photoluminescence quantum yield (PLQY) and time-resolved photoluminescence (TRPL) measurements for both pristine-PeNPL and BPAc-PeNPL colloidal solutions. As shown in **Supplementary Fig. 12**, BPAc-PeNPLs exhibit a maximum PLQY of 65.7% and a median PLQY of 62.7 ± 2.8% across 9 independently prepared samples. In contrast, pristine-PeNPLs show significantly lower values, with a maximum PLQY of 36.1% and a median value of 34.3 ± 2.2% across 9 independently-prepared samples. The increased PLQY of BPAc-PeNPLs suggests effective passivation of surface defects, reducing nonradiative recombination. Time-resolved PL decay measurements further support the strongly passivating nature of BPAc ligands. Across all excitation fluences, the average carrier lifetime in BPAc-PeNPLs exceeds that in pristine-PeNPLs, consistent with reduced non-radiative recombination (**Supplementary Fig. 13, Supplementary Table 1**). These findings highlight the critical role of BPAc in suppressing defect-induced nonradiative losses, resulting in improved optical properties of PeNPLs. This can also be supported by the decay curves extracted from transient absorption (TA) spectroscopy measurements of pristine- and BPAc-PeNPL colloidal solutions shown in **Supplementary Fig. 14**, where the pristine-PeNPLs exhibited faster decay than BPAc-PeNP Ls. Experimental evidence for BPAc coordination is also observed in $^{1}$H-NMR measurements (detailed discussion in **Supplementary Note 1**). These experimental observations are consistent with the DFT-predicted strong binding between BPAc ligand and perovskite surface, reinforcing the dual role of BPAc ancillary ligand as both a kinetic modulator and an effective surface passivator. This dual functionality leads to the formation of highly uniform 3ML $CsPbI_3$ PeNPLs with enhanced optical properties, including suppressed nonradiative recombination and prolonged charge-carrier lifetimes. Such improvements in structural and photophysical properties are critical prerequisites for the formation of well-ordered, and highly-emissive PeNPL superlattices, which will be discussed in the following section.

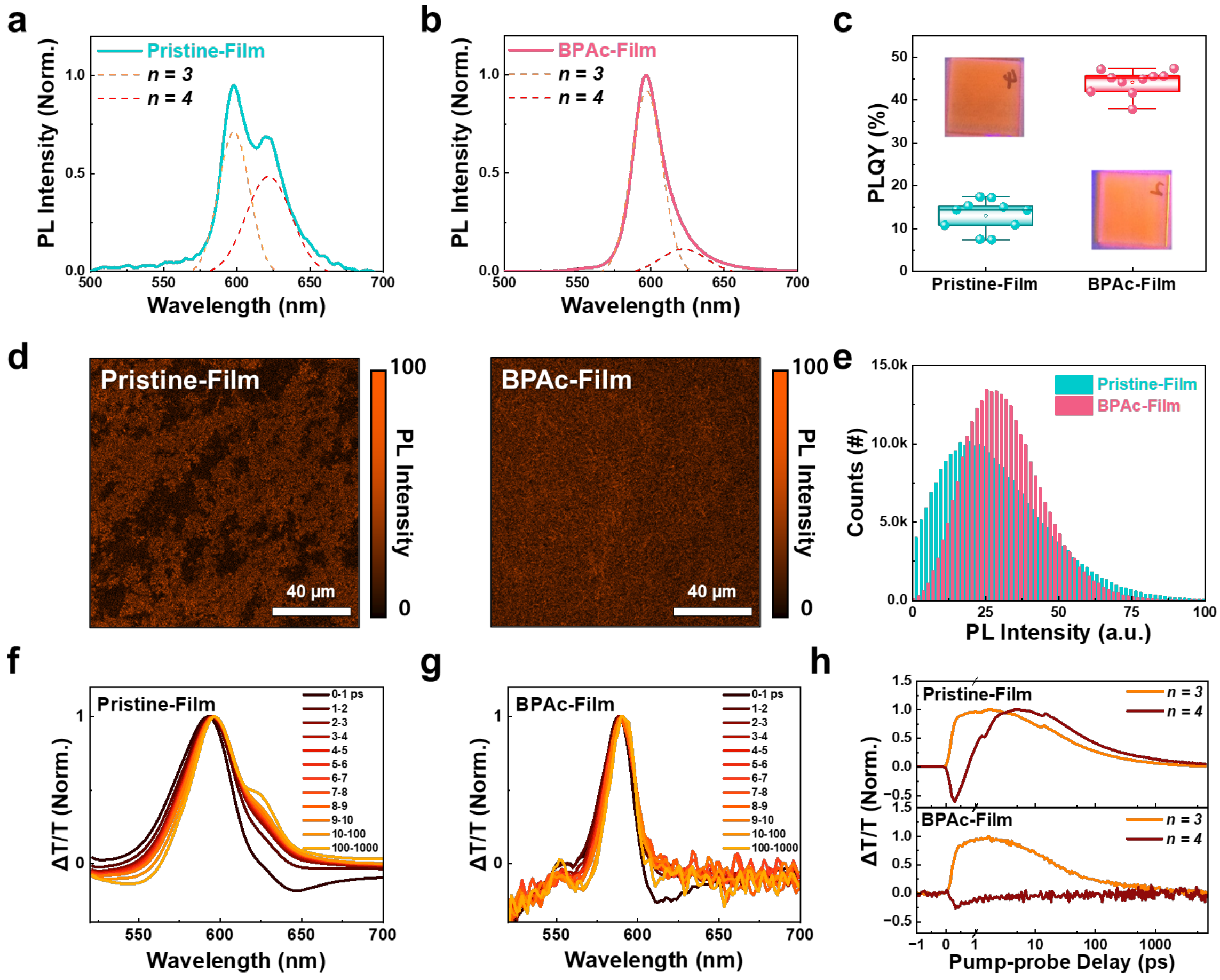

**Fig. 2 | Optical characterization of pristine- and BPA-treated $CsPbI_3$ PeNPL films**

PL spectra of **a** Pristine-PeNPL thin films and **b** BPAc-PeNPL thin films showing contributions from 3ML and 4ML components. **c** PLQY distributions and **d** confocal PL images of each PeNPL thin film (scan area: 120 × 120 μm$^2$). **e** PL intensity histograms extracted from the confocal PL mapping images. Normalized TA spectra from **f** pristine- and **g** BPAc-PeNPL films over the first 1000 ps after photoexcitation, excited with a 400 nm wavelength laser (17.28 μJ cm$^{-2}$ fluence). **h** Charge-carrier kinetics of the 3 ML (probed at 580 – 600 nm wavelength) and 4 ML (probed at 610–630 nm wavelength) PeNPLs present in the films. Positive signals represent the ground states bleach (GSB) signals. $\Delta T/T$ values normalized to the peak GSB for each trace.

## Perovskite nanoplatelet uniformity in films

Depositing PeNPLs from colloidal solution to form films can lead to ligand detachment and PeNPL agglomeration. To investigate the film-formation behavior of PeNPLs, each PeNPL thin film was fabricated by spin coating the colloidal solution. Indeed, pristine-PeNPL thin films (**Fig. 2a**) show broadened emission with two distinct peaks at 600 nm and 627 nm, corresponding to 3ML and 4ML PeNPLs, respectively. This spectral broadening is more pronounced than that observed in colloidal solution (**Fig. 1f**), indicative of NPL agglomeration[45]. In contrast, BPAc-PeNPL thin films (**Fig. 2b**) exhibit a dominant, narrow PL peak centered at 600 nm. Spectral deconvolution reveals that the 600 nm component accounts for 53.6% of the total PL in pristine films, increasing to 84.7% in BPAc-PeNPL films, with the 627 nm wavelength component of the total PL emission reduced from 46.4% to 15.3%. The more dominant PL emission at 600 nm (3ML) can be attributed to the strong coordination between BPAc ligands and the perovskite surface, which prevents aggregation and preserves the size distribution even during thin film formation. PLQY measurements of the PeNPL films further supports these observations. As shown in **Fig. 2c**, pristine-PeNPL thin films exhibit a PLQY of 17.4% (median: 13.0% across 9 samples). In contrast, BPAc-PeNPL thin films retain a high PLQY of 47.4% (median: 44.0% across 9 samples), comparable with the PLQY in colloidal solution. This result is consistent with the passivation effects observed in the solution-phase measurements and confirms that BPAc ligand binding effectively suppresses nonradiative recombination pathways even in thin film form.

The uniformity of the PeNPL film was evaluated through PL mapping using a confocal microscope (**Fig. 2d**), with the corresponding intensity histograms shown in **Fig. 2e**. Pristine-PeNPL thin films display heterogeneous emission intensity across the scanned area, and has a broad distribution in PL intensities (**Fig. 2e**). In contrast, BPAc-PeNPL thin films exhibit significantly enhanced PL uniformity, with narrower and more symmetric intensity profiles. This homogeneity is attributed to the enhanced PeNPL monodispersity and strong surface passivation provided by the BPAc ancillary ligand.

The improved monodispersity also influences the charge-carrier dynamics in PeNPL thin films. Although the presence of ligands can suppress charge transfer between different PeNPLs[46], energy transfer between PeNPLs with different thicknesses (and thus different optical gaps) can still occur and degrade device performance[47, 48]. This energy transfer between NPLs of different thicknesses is much less pronounced in colloidal solution, where the NPLs are relatively far apart and separated by solvent molecules. **Supplementary Fig. 14a** shows the normalized TA spectra of the pristine PeNPLs in solution. Both characteristic ground state bleach (GSB) peaks of the 3ML (580–600 nm) and 4ML (610–630 nm) populations are present. The relative intensity ratio between these two peaks does not change over time, and the GSB decay curves for the two thicknesses show no overlap at early times ($< 1$ ns), indicating negligible energy transfer between NPLs in solution. In contrast, **Fig. 2f** presents the normalized TA spectra of the pristine PeNPL thin film, where the relative intensity of the 4ML bleach peak initially increases over time. Moreover, **Fig. 2h** shows clear overlap between the GSB decay curves of the 3ML and 4ML PeNPLs within each film. Together, these observations suggest substantial energy transfer in the pristine samples from the higher optical-gap 3ML PeNPLs to the lower optical-gap 4ML PeNPLs in the film. This additional energy-transfer pathway introduces an extra nonradiative decay channel for photoexcited or electrically-injected charge-carriers, thereby reducing overall device performance and compromising emission purity. By comparison, the BPAc-PeNPLs exhibit a more uniform thickness distribution. The substantially suppressed population of 4ML PeNPLs results in the 4ML GSB peak at 610–630 nm being negligible above noise in both the solution and film samples, as shown in **Supplementary Fig. 14b** and **Fig. 2g**. This suppression of inter-thickness energy transfer enables more efficient radiative recombination in BPAc-PeNPL thin films.

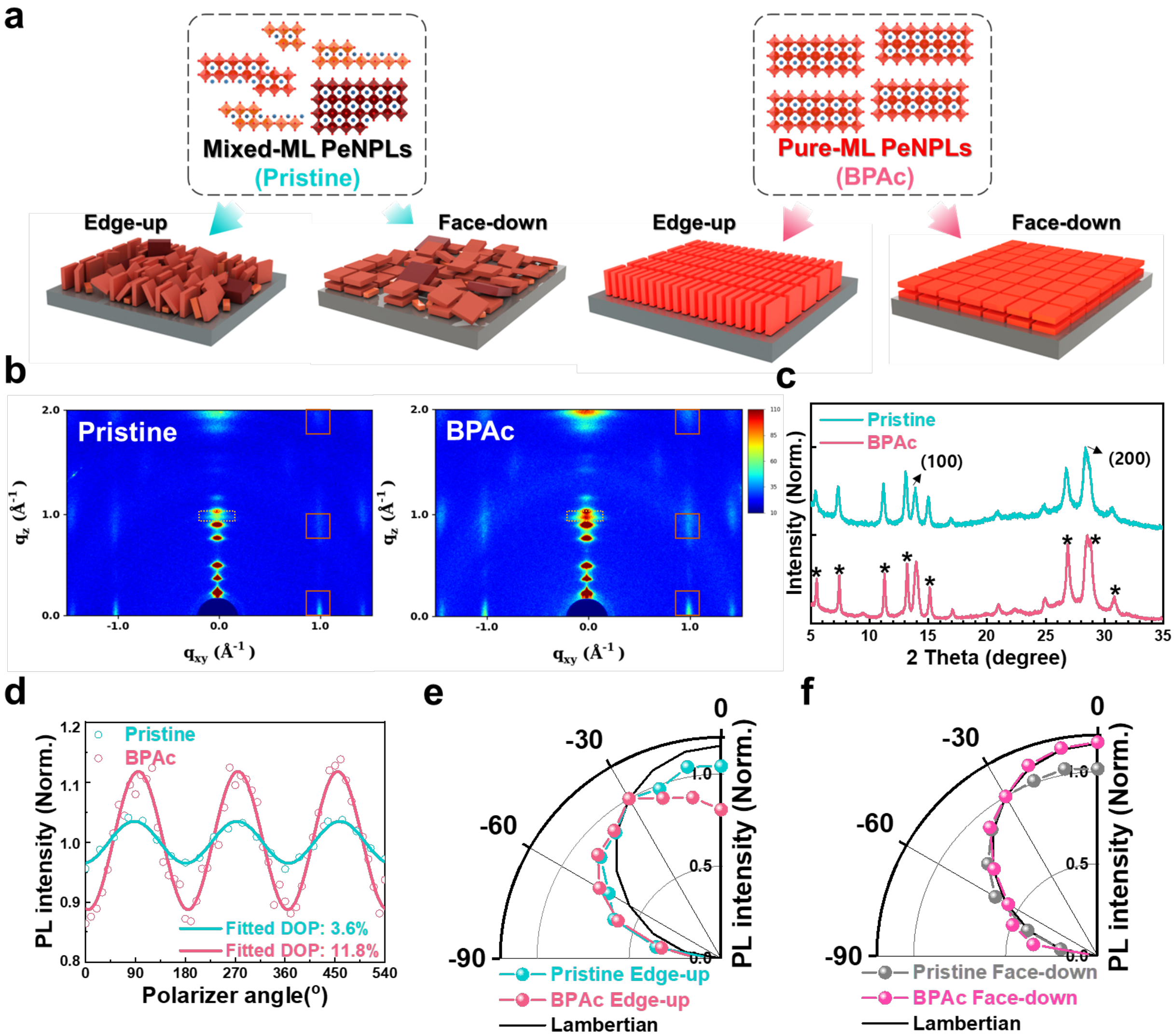

**Fig. 3 | Structural characterization and linearly polarized light emission from superlattice films**

**a** Schematic illustration of superlattice formation from mixed-ML and uniform-ML PeNPLs. **b** Grazing-incidence wide-angle X-ray scattering (GIWAXS) patterns and **c** X-ray diffraction (XRD) patterns of face-down oriented pristine- and BPAc-PeNPLs. The peaks marked with stars correspond to the superlattice peaks, while the Bragg peaks are labelled with Miller indices. **d** Polarization angle dependence of normalized PL intensity of pristine- and BPAc-PeNPL thin films, fitted with a sinusoidal function. Lambertian profile and angular distribution of the normalized PL intensity from **e** edge-up and **f** face-down pristine- and BPAc-PeNPL thin films.

## Formation and characterization of PeNPL superlattices

To further evaluate the structural properties of the synthesized PeNPLs, we investigated their formation into superlattices. The orientation of the PeNPLs can be tuned from edge-up to face-down by controlling the evaporation rate of the alkane solvent used to disperse the PeNPLs [18, 42, 49]. A schematic illustration of the self-assembly of PeNPLs is shown in **Fig. 3a**, where BPAc-PeNPLs, due to their narrow size and thickness distribution, form highly ordered superlattices. In contrast, pristine-PeNPLs, with broader size and shape variations, exhibit disordered and randomly stacked superlattices. To characterize the structural orientation and periodicity of the PeNPL superlattices, grazing-incidence wide-angle X-ray scattering (GIWAXS) measurements were conducted (**Fig. 3b** and **Supplementary Fig. 18**). Each PeNPL thin film was fabricated by spin coating colloidal PeNPL dispersions in octane and hexane, which preferentially promoted face-down and edge-up orientations, respectively. As a result, BPAc-PeNPL thin films exhibited a series of well-defined out-of-plane diffraction peaks corresponding to the (100), (101), and (102) planes of the perovskite structure, which were markedly sharper and more intense than those of the pristine-PeNPL thin films. The vertical spread of the diffraction features further confirms coherent layer-by-layer stacking in the BPAc-PeNPL films. One-dimensional scattering profiles extracted from the integration of diffraction rings (**Supplementary Fig. 19**) show low-angle peaks ($q < 1$ Å$^{-1}$), which can be attributed to periodic electron density variations between the dense inorganic slabs and the surrounding organic ligands[18, 21]. These features were especially pronounced in edge-up oriented BPAc-PeNPL thin films, supporting the formation of superlattices with consistent interlayer spacing. To quantify the degree of orientational alignment, we further analyzed the azimuthal dependence of scattering intensity (**Supplementary Fig. 20**). In the BPAc-PeNPL thin films, face-down orientations exhibited stronger intensity in the low azimuthal angle range (0°–10°), while edge-up orientations showed enhanced scattering at higher angles (80°–90°). These patterns were significantly sharper than those of pristine samples, indicating a higher degree of orientation in both face-down and edge-up PeNPL thin films due to the improved monodispersity of BPAc-PeNPLs that allowed more regular stacking. Finally, powder X-ray diffraction (XRD) patterns were recorded to confirm the periodic stacking behavior (**Fig. 3c** and **Supplementary Figs. 21-22**). Both PeNPL thin films exhibited multiple satellite peaks in addition to the Bragg peaks, in agreement with prior reports on PeNPL superlattices[49, 50]. Notably, BPAc-PeNPL thin films showed narrower FWHM values for these satellite peaks, indicating superior long-range structural coherence and crystallinity. Collectively, these results highlight the critical role of the BPAc ancillary ligand in enabling the formation of highly-ordered, structurally-coherent PeNPL superlattices. As demonstrated in the preceding analyses, the well-orientated PeNPL superlattices provide a versatile platform for a variety of applications. Edge-up oriented PeNPL thin films, characterized by their high degree of in-plane dipole alignment, are advantageous for achieving strong linear polarization in light emission. In contrast, face-down oriented PeNPL thin films, with their efficient out-coupling, are suitable for LED devices, where maximizing light extraction is essential[51, 52, 53, 54]. These orientation-dependent properties will be further explored in the following sections, demonstrating the broad applicability of PeNPL superlattices in both polarization-sensitive photonic applications and efficient light-emitting diodes (LEDs).

## Linearly polarized light emission and improved out-coupling

Ultrathin PeNPL light emitters inherently exhibit linearly polarized light emission at the single particle level through exciton fine structure splitting. But maintaining a high degree of polarization at the device level requires fine control over the orientation and alignment of the PeNPLs[55, 56]. Previously, we established that a high degree of polarization (DOP) in emission from the normal direction to the substrate requires edge-up PeNPLs[18, 25, 42, 54]. We therefore focused on edge-up PeNPL superlattice films, and measured the DOP, as detailed in **Supplementary Fig. 23**. As shown in **Fig. 3d**, Pristine-PeNPL films exhibited a relatively low DOP, with a maximum of 3.6% and a median of 3.4% across 5 independently prepared samples. In contrast, BPAc-PeNPL thin films displayed a significantly enhanced polarization response, with a maximum DOP of 11.8% and a median of 11.5% across 5 independently prepared samples. The corresponding DOP histogram is provided in **Supplementary Fig. 24-25**, with individual data points summarized. Angle-resolved PL measurements of edge-up oriented PeNPL superlattice films further supports this trend (**Fig. 3e**). In these measurements, 0° corresponds to light collection normal to the thin film surface, while 90° represents collection parallel to the thin film plane. The PL intensity variation with respect to collection angle was significantly more pronounced in BPAc-PeNPL thin films than in their pristine-PeNPL thin films. This behaviour indicates enhanced transition dipole alignment, as evidenced by the GIWAXS results, which show that an increase in the percentage of PeNPLs oriented edge-up (**Supplementary Fig. 20**) from pristine (50%) to BPAc-PeNPLs (54.5%).

The advantage of being able to tune the orientation of PeNPL film is that we can also tune their orientation to face-down in order to maximize outcoupling in the normal direction by maximizing the fraction of horizontal transition dipole moments[54, 57, 58]. The angular distribution profiles of PL from the face-down superlattices based on pristine-PeNPLs and BPAc-PeNPLs is shown in **Fig. 3f**. These measurements also show that the BPAc-PeNPLs exhibit a more Lambertian profile, in contrast to the more directional emission observed in face-down oriented pristine-PeNPL thin films. The improved angular emission profile with BPAc treatment is consistent with our previous optical simulations of $CsPbI_3$ PeNPLs indicating that a higher fraction of horizontal transition dipole moments leads to more Lambertian-like emission[18], in line with the more pronounced face-down alignment suggested by our GIWAXS results (**Supplementary Fig. 20c**). These results also suggest that BPAc-PeNPLs would exhibit higher outcoupling, which is particularly advantageous for LED applications[51, 52, 59].

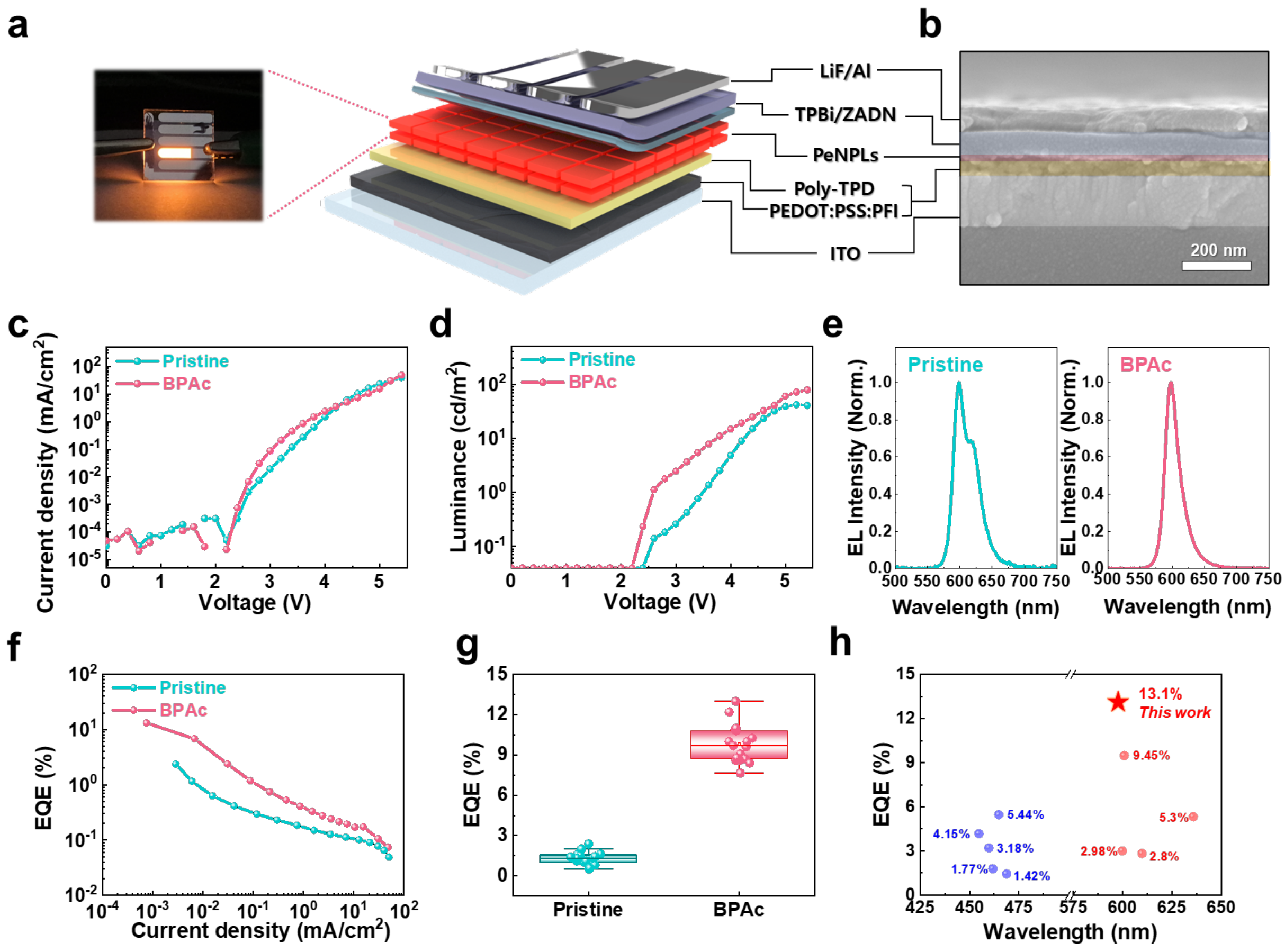

**Fig. 4 | Device performance of pristine- and BPA-treated $CsPbI_3$ PeNPLs**

**a** Device architecture and **b** cross sectional scanning electron microscopy (SEM) images of PeNPL LED devices. Comparison of the **c** current density–voltage (*J–V*) curve, **d** luminance–voltage (*L–V*) curve, **e** electroluminescence (EL) spectra, and **f** external quantum efficiency (EQE)–current density (*EQE–J*) curves of champion devices. **g** $EQE_{max}$ histogram from 17 individual PeNPL devices of each type. **h** Comparison of the $EQE_{max}$ of strongly-confined $CsPbX_3$ (X = I, Br) PeNPL LEDs in the ultrathin regime ($n \leq 3$) with previous literature.[18, 28, 29, 60, 61, 62, 63, 64, 65]

**Table 1. Device performance of champion LEDs based on pristine- and BPAc-PeNPLs**

| Sample configuration | $EQE_{max}$ [%] @ bias | $L_{max}$ [cd/m²] @ bias | $LE_{max}$ [cd/A] @ bias | Turn-on Voltage [V] @ 0.1 cd/m² |
|---|---|---|---|---|
| Pristine-PeNPLs | 2.4@2.6V | 42@4.8V | 5.1@2.6V | 2.6 |

| BPAc-PeNPLs | 13.1@2.4V | 78@5.4V | 31.2@2.4V | 2.4 |
|---|---|---|---|---|

## Realizing efficient strongly-confined PeNPL LEDs in the ultrathin regime

To evaluate the optoelectronic performance of the synthesized PeNPLs, we fabricated LEDs with the device architecture illustrated in **Fig. 4a**. The device stack consists of an indium tin oxide (ITO) transparent electrode, followed by a bilayer hole transport layer composed of PEDOT:PSS:PFI and poly(N,N′-bis-4-butylphenyl-N,N′-bisphenyl)benzidine (Poly-TPD). The emissive layer comprises of either pristine- or BPAc-PeNPLs. To maximize outcoupling, we spin-coated the PeNPLs from a colloidal solution in octane to achieve a face-down orientation. A bilayer electron transport structure comprised of 2,2′,2″-(1,3,5-benzinetriyl)-tris(1-phenyl-1-H-benzimidazole) (TPBi) and 2-[4-(9,10-di-naphthalen-2-yl-anthracen-2-yl)-phenyl]-1-phenyl-1H-benzoimidazole (ZADN) was deposited on top of the PeNPL layer, followed by LiF/Al as the top electrode. A cross-sectional scanning electron microscopy (SEM) image of the complete device structure is shown in **Fig. 4b**, confirming uniform layer stacking and a smooth interface with the PeNPL emissive layer. The orientation of PeNPLs within the thin film significantly influences device performance due to the optical out-coupling effect. Among the two orientations investigated in the previous section (edge-up and face-down orientation), the face-down orientation was selected for the primary device architecture due to its superior outcoupling in the normal direction. The device performance results of edge-up oriented LEDs are provided in **Supplementary Fig. 26**, which exhibit lower efficiencies than the face-down PeNPL devices due to reduced outcoupling in the normal direction. This performance difference highlights the critical role of orientation in optimizing light extraction and achieving high-efficiency PeNPL-based LEDs. As shown in the current density–voltage (*J*–*V*) and luminance–voltage (*L*–*V*) characteristics (**Fig. 4c** and **4d**), devices based on BPAc-PeNPLs exhibit improved charge-carrier injection and a lower turn-on voltage compared to pristine-PeNPL devices. This improvement is attributed to the partial replacement of long-chain, insulating native ligands in pristine-PeNPLs with shorter BPAc ligands, which enhance charge carrier mobility and facilitate stronger interparticle coupling within the emissive layer. Additionally, the well-ordered superlattice formed by the uniform-ML BPAc-PeNPLs promotes efficient charge transport and a uniform and narrow emission profile, as discussed in the section on PeNPL superlattice structural characterization. Electroluminescence (EL) spectra of the pristine-PeNPL and BPAc-PeNPL LED devices are shown in **Fig. 4e**. Pristine-PeNPL LEDs exhibited broad EL profiles with multiple emission peaks corresponding to mixed 3ML and 4ML PeNPLs, consistent with PL measurements from these films (**Fig. 2a**). In contrast, BPAc-PeNPL LEDs displayed a single, narrow EL peak centered at 600 nm wavelength, attributed to uniform 3ML PeNPL emission. This spectral purity, resulting from the improved well-controlled thickness achieved through BPAc incorporation, is further corroborated by the absence of multiple EL peaks in BPAc-PeNPL devices, confirming the structural integrity and strong exciton confinement of the PeNPL layer. The champion BPAc-PeNPL device achieved a maximum EQE of 13.1%, significantly surpassing the 2.4% EQE of the champion pristine-PeNPL device (**Fig. 4f** and **Table 1**). This performance was further validated by statistical analysis of 17 independently-fabricated devices (**Fig. 4g**), demonstrating high reproducibility and a narrow distribution in EQE values. To the best of our knowledge, this value represents the highest reported EQE for strongly confined colloidal $CsPbX_3$ PeNPL-based LEDs in the ultrathin regime (*i.e.*, 3 ML thick or thinner, **Fig. 4h** and **Supplementary Fig. 27**). As explained in the introduction, this ultrathin regime is critical for realizing polarized light emission and is much more challenging to achieve efficient device performance due to the prominent role of surfaces, and greater difficulty in controlling thickness uniformity. The enhanced device performance observed in BPAc-PeNPL LEDs can be attributed to several key factors. First is the efficient outcoupling arising from the well-aligned, face-down PeNPL superlattices with precisely controlled shape and thickness (**Fig. 3f**). Secondly, the strong binding of the ancillary BPAc ligands to the perovskite surface effectively passivates surface defects, reducing nonradiative losses. These synergistic effects contribute to the superior device performance of BPAc-PeNPL LED devices compared to the pristine-PeNPL LEDs. Overall, the successful integration of BPAc as an ancillary ligand not only enhances the device efficiency of PeNPL-based LEDs but also improves their linear polarization properties. These findings demonstrate the feasibility of ligand engineering as a promising approach to advance the development of high-efficiency, strongly confined PeNPL-based LEDs with potential applications in both high-performance LEDs and linearly polarized quantum light sources.

## DISCUSSION

This work reveals how the molecular structure of the ancillary ligand introduced during synthesis governs the nucleation, growth, and assembly of $CsPbI_3$ nanoplatelets through a synergistic balance between headgroup coordination and backbone steric effects. The coordination strength of the functional group to $Pb^{2+}$ influences the barrier to nanoplatelet nucleation, while the steric profile of the organic ligand backbone dictates anisotropic growth and colloidal stability. When these two factors are optimally balanced, the growth pathway converges toward uniform monolayer PeNPLs with well-passivated surfaces, providing a molecular rationale that links ligand chemistry to nanoscale morphology and photonic function. The resulting PeNPLs exhibit enhanced optical properties, suppressed nonradiative losses, and tunable optical anisotropy. By precisely controlling the orientation of the PeNPLs, we achieved a remarkable EQE of 13.1% in LED devices through improved out-coupling in face-down oriented superlattices, whereas edge-up oriented superlattices exhibit enhanced linearly polarized light emission. Overall, our findings show that achieving PeNPLs with improved uniformity and PLQY can be realized using ancillary ligands containing a strong Lewis base in the functional group, as well as an organic backbone that is not overly bulky, such that they do not displace native ligands on the PeNPL surface. These design principles can be more broadly applied across the wider PeNPL family, as well as other NPL materials systems. The improvements in performance and uniformity can allow the multifunctionality of anisotropic NPLs to be fully utilized, particularly NPLs in the more challenging ultrathin regime required for achieving polarized light emission.

## DATA AVAILABILITY

Raw data for the main text and supplementary information available from the Oxford Research Archive repository [DOI to be made available before publication]

## CONFLICT OF INTERSET

The authors declare no competing interests.

## ACKNOWLEDGEMENTS

J. Y. and R. L. Z. H. thank the UK Research and Innovation for funding through a Frontier Grant (no. EP/X029900/1), awarded via the 2021 ERC Starting Grant scheme. J. Y. and R. L. Z. H. also thank St. John's College Oxford for funding through the Welcome and Large Grant schemes. R. L. Z. H. thanks the Royal Academy of Engineering and Science & Technology Facilities Council for funding through a Senior Research Fellowship (no. RCSRF2324-18-68). The authors thank the Engineering & Physical Sciences Research Council (EPSRC; no. EP/V061747/1) for support. Y. W. thanks Merton College Oxfrod, as well as the Department of Chemistry at the University of Oxford for financial support. C.-Y. C. acknowledges support from the Oxford-Taiwan Graduate scholarship, as well as from the Clarendon Fund. This work was supported by the National Research Foundation of Korea (NRF) grant funded by the Korea government (MSIT) (RS-2025-00514448 and RS-2025-00516815) and contains the results obtained by using the equipment of UNIST Central Research Facilities (UCRF). M.S.I. and A.N.A. acknowledge the Department of Materials at the University of Oxford for a DTP DPhil studentship. M.S.I. and V. acknowledge the EPSRC Programme Grant (EP/X038777/1) for a postdoctoral fellowship. M.S.I, A.N.A. and V are grateful to the UK's HEC Materials Chemistry Consortium (EP/X035859/1) for ARCHER2 high-performance computing facilities.

## AUTHOR CONTRIBUTION

J.K. and W.H.J. contributed equally to this work. J.K., W.H.J., J.Y. and R.L.Z.H. conceived of the ideas behind this work. J.K., W.H.J. performed formal analysis under the supervision of M.H.S., B.R.L. and R.L.Z.H., A.N.A and V. performed DFT calculations under the supervision of M.S.I. D.K. and D.L performed confocal PL microscopy. Y.-T.H. and Y.W. performed transient absorption spectroscopy under the supervision of A.R. C.-Y.C. and J.K performed the TEM measurements. X.S. and H.J.S. performed optical analysis, S.Y.B. performed XRD. J.K. and W.H.J. drafted the first version of the manuscript. J.Y., M.S.I., B.R.L., M.S. and R.L.Z.H. reviewed and revised the manuscript. This work was carried out under supervision of B.R.L., M.H.S and R.L.Z.H. All the authors contributed to the discussion of the manuscript.